\documentclass[11pt,citesort]{article}

\usepackage{geometry}
\usepackage{axodraw}
\newcommand{\ice}[1]{\relax}
\usepackage{graphicx}
\usepackage{epsfig}
\usepackage{latexsym}
\usepackage{amsmath}
\usepackage{amsfonts}
\usepackage{amsxtra}
\usepackage{cite}

\def\krto{ {\,\,\lower .8ex\hbox {$\longrightarrow \atop k \rightarrow 0$}\,\,}}
\def\Section#1{\section{#1}\hspace{\parindent}}

\def\bea{\begin{eqnarray} }
\def\beq{\begin{eqnarray} }

\def\eea{\end{eqnarray}}
\def\eeq{\end{eqnarray}}
\def\eq#1{eq.~(\ref{#1})}
\def\Eq#1{Eq.~(\ref{#1})}
\def\eqs#1{eqs.~(\ref{#1})}

\begin{document} 
\date{}

\title{ Gribov's horizon and the ghost dressing function}
\author{ 
Ph.~Boucaud$^a$, 
J.P.~Leroy$^a$, 
A.~Le~Yaouanc$^a$,
J. Micheli$^a$, \\
O. P\`ene$^a$, 
J.~Rodr\'iguez-Quintero$^b$
}

\maketitle

\begin{center}
$^a$Laboratoire de Physique Th\'eorique\footnote{Unit\'e Mixte de Recherche 8627 du Centre National de 
la Recherche Scientifique} \\
Universit\'e de Paris XI, B\^atiment 210, 91405 Orsay Cedex,
France \\
$^b$ Dpto. F\'isica Aplicada, Fac. Ciencias Experimentales,\\
Universidad de Huelva, 21071 Huelva, Spain.
\end{center}

\begin{abstract}
We study a relation recently derived by K. Kondo at zero  momentum between the
Zwanziger's horizon function, the ghost dressing function and Kugo's functions $u$ and $w$.  We agree with this result as far as bare quantities are considered. 
However, assuming the validity of the horizon gap equation, we argue that the
solution $w(0)=0$ is not acceptable since it  would lead to a vanishing
renormalised ghost dressing function. On the contrary, when the cut-off goes to infinity, 
$u(0) \to \infty$, $w(0) \to -\infty$ such that  $u(0)+w(0) \to -1$.
Furthermore $w$ and $u$ are not multiplicatively renormalisable. Relaxing 
the gap equation allows $w(0)=0$ with $u(0) \to -1$. In both cases the bare
ghost dressing function, $F(0,\Lambda)$, goes logarithmically to infinity at 
infinite cut-off.  We show that, 
although the lattice results provide bare results not so  different from the 
$F(0,\Lambda)=3$ solution, this is an accident due to the fact that the lattice
cut-offs lie in the range 1-3 GeV$^{-1}$. We show that the renormalised ghost
dressing function should be finite and  non-zero at zero momentum and can be
reliably estimated on the lattice  up to powers of the lattice spacing~; from published
data on a  $80^4$ lattice at $\beta=5.7$ we obtain $F_R(0,\mu=1.5$  GeV)$\simeq
2.2$. 
\end{abstract}

\begin{flushright}
{\small UHU-FP/09-019}\\
{\small LPT-Orsay/09-62}\\
\end{flushright}




\Section{Introduction}

Kondo has derived in recent papers~\cite{Kondo:2009ug,Kondo:2009gc}
a relation between the $k=0$ values of the  ghost dressing function $F(k)$, Zwanziger's horizon 
function $h(k)$,  Kugo's function $u(k)$, and an additional function
$w(k)$. Applying to this relation Zwanziger's horizon gap equation and 
assuming that $w(0)=0$  he derives the surprising result that $u(0)=-2/3$.
This is surprising as so simple constraints on bare quantities are rare. 
We know only the case of the electric charge which benefits of the Ward identity.
This surprising result deserves some closer investigation, even more so as 
lattice results are not so far from it as we shall see. Indeed it has given
rise to several publications and there is far from a consensus on this matter 
\cite{Zwanziger:2009je, Dudal:2008sp, Dudal:2009xh, Maas:2008ri}.

To understand better the issue we try in this note to reconsider 
every point of the discussion from first principles. Our
starting point is a set of relations between the functions we have just mentionned.
 They
 concern bare quantities, which imposes to use a finite ultraviolet cut-off, 
 else we would have to deal with divergent quantities. 
 In section  \ref{sec2} we propose a faster derivation of these relations.
 If one assumes the validity of the Zwanziger horizon gap equation this 
 boils down to very simple relations giving $u(0)$ and $w(0)$ as  functions of 
 $F(0)$. 
 We then discuss whether $u(0)$ can be -2/3 or any finite quantity. We argue that it
 is not possible if we assume $F$ to be multiplicatively renormalisable, which
 nobody would deny. 
 In section \ref{sec3} we use lattice QCD and  ghost propagator
 Dyson-Swinger equation (GPDSeq) to get numbers.     From the GPDSeq at small 
 momentum we find that ratio of the the bare (resp. renormalised) ghost dressing funtions at small and zero momentum, assuming
 the latter  to be finite, is  essentially cut-off and renormalisation point independent. 
 We extract an estimate of the renormalised $F_R(0)$. 
 Finally we discuss the status of the Zwanziger horizon gap equation on the
 lattice. Convinced that it has no reason to be valid, we generalise the result
 of section \ref{sec2} for a more general case. 
   
\section{Ghost dressing function, horizon function, $u$ and $w$}
\label{sec2}

The discussion which follows  deals with bare quantities. These 
are singular and need a regulator, or cut-off, which we will call $\Lambda$
(in the lattice case, this regulator is $a^{-1}$, $a$ being the lattice spacing).
The dependence in $\Lambda$ will often be kept implicit, to avoid heavy 
notations, but is always understood speaking of bare quantities. 
Renormalised quantities will be marked by the index $R$. There is no need to  specify the   renormalisation scheme being used, since our results do not depend on a particular choice; however, regarding lattice results, we shall refer as usual to the MOM scheme. 
\subsection{Gribov-Zwanziger action}
In~\cite{Kondo:2009ug,Kondo:2009gc}, it has been
claimed that  three-point and four-point functions for gluon and ghost fields
can be related in such a manner that the Zwanziger horizon condition  strongly
constrains the ghost propagator and the ghost-gluon vertex.

It is well known, since Gribov's famous paper~\cite{Gribov:1977wm}, that 
the gauge fixing procedure in QCD using the standard Faddeev-Popov procedure is not unambiguous.
It leads to a discrete set of solutions, named ``Gribov copies''. 
One solution, 
 proposed by Zwanziger~\cite{Zwanziger:1989mf} {\it ,
which aims at restricting the Gribov copies within the Gribov Horizon}, consists in using
the Gribov-Zwanziger partition function in Landau gauge,

\beq\label{GZ} Z_\gamma = \int \left[ D A \right]  \delta\left( \partial A \right) \ {\rm
det}(M) \ e^{-S_{\rm YM} + \ \gamma \int d^Dx h(x)} \ , \eeq
for the D-dimensional Euclidean Yang-Mills theory, 
where $S_{\rm YM}$ stands for the Yang-Mills action, $M$ is the 
Faddeev-Popov operator, 
\beq
M^{ab} = - \partial_\mu D^{ab}_{\mu} = - \partial_\mu \left( \partial_{\mu} 
\delta^{ab} + g f^{abc} A_\mu^c \right) 
\eeq
and $h(x)$ is the Zwanziger horizon function,
\beq
h(x)=  \int d^Dy \ g f^{abc} A_\mu^b(x) (M^{-1})^{ce}(x,y) g f^{afe}A_\mu^f(y) \ ;
\eeq 
that restricts the integration over the gauge group to the first Gribov region, provided 
that the Gribov parameter, $\gamma$, is a positive number 
that is to be determined by solving the 
so-called gap equation:
\beq\label{horizon1}
\langle h(x) \rangle_\gamma = \left( N^2 -1 \right) \ D \ .
\eeq  
%
%
The horizon function is a bare quantity depending on the cut-off parameter, as
$\gamma$  also does through the implementation of the horizon condition that
requires that the gap equation  be solved for 	every cut-off value.
We will postpone the discussion of this gap equation to a later section. In any case, the horizon 
function is a well defined bare quantity and it is relevant to first derive,
{\it independently of \eq{horizon1}} the relation its v.e.v. has with other 
quantities.

\subsection{Relating $h(0), u(0), w(0)$ and $F(0)$}

In this subsection we propose a simplified derivation of Kondo's
 relation (cf~\cite{Kondo:2009ug,Kondo:2009gc}) which relates
the v.e.v. of the horizon function $h(0)$ to
 the ghost dressing function at vanishing momentum and the Kugo-Ojima 
 parameters. Then, contrarily to Kondo, we will add no assumption 
about the Kugo-Ojima parameters but simply combine Kondo's relation  with the one discovered by  
Kugo between the ghost dressing in Landau gauge with these Kugo-Ojima
parameters and discuss about their general implications.
\beq\label{horizon2}
\langle h(0) \rangle_{k=0} &=& 
 \lim_{k^2 \to 0}
\frac 1 {V_D} \int d^Dx \int d^D y \ 
\langle g f^{abc} A_\mu^b(x) (M^{-1})^{ce}(x,y) g f^{afe} A_\mu^f(y) \rangle 
\ e^{i k \cdot (x-y)} \nonumber \\
&=&
 \lim_{k \to 0}
\frac 1 {V_D} \int d^Dx \int d^D y \ 
\langle \ \left( g f^{abc} A_\mu^b \ c^c\right)_{x}   
\left( g f^{afe} A_\mu^f \ \overline{c}^e \right)_{y} \ \rangle 
e^{i k \cdot (x-y)} \nonumber \\
&=&
 \lim_{k^2 \to 0}
\int d^D (x-y) \ 
\langle \ \left( g f^{abc} A_\mu^b \ c^c\right)_{x}   
\left( g f^{afe} A_\mu^f \ \overline{c}^e \right)_{y} \ \rangle 
e^{i k \cdot (x-y)} \nonumber \\ 
&=&
 \langle \ \left( g f^{abc} A_\mu^b \ c^c\right)   
\left( g f^{afe} A_\mu^f \ \overline{c}^e \right) \ \rangle_{k^2 \to 0} 
\eeq 
where we use the simplified notation:
\beq
\langle \ (\dots) (\dots) \ \rangle_k \ 
\equiv \ \int d^D(x-y) \langle (\dots)_x (\dots)_y \rangle \ e^{ i k \cdot (x-y)} \  
\eeq
that was introduced in ref.~\cite{Kondo:2009ug,Kondo:2009gc} and that will be followed 
from now on. To establish \eq{horizon2}, nothing is needed but the relation between the inverse Faddeev-Popov operator and the ghost and anti-ghost fields 
%
%
and the translational invariance.
Define  then 
the function $u(k^2)$, the  value of which at  vanishing momentum gives the Kugo-Ojima parameter, as~
\beq\label{KOpar2}
\langle \ \left(D^{ab}_\mu c^b \right) \left(g f^{cde} A_\nu^d \overline{c}^e \right) \ \rangle_k
\ = -\ 
\delta^{T}_{\mu\nu}\delta^{ac} \ u(k^2) \ ;
\eeq
where $k^2 \delta^{T}_{\mu\nu}(k)\equiv k^2 \delta_{\mu\nu} - k_\mu k_\nu$ and 
the transversality is guaranteed by the well-known identity:
\beq\label{ident0}
\langle \ \left(\partial_\mu D^{ab}_\mu c^b \right) \left(g f^{cde} A_\nu^d \overline{c}^e \right) \ \rangle_k
\ = \ 
-i k_\mu \langle \ \left(D^{ab}_\mu c^b \right) \left(g f^{cde} A_\nu^d \overline{c}^e \right) \ \rangle_k
\ = \ 0 \ .
\eeq
Now, by merely invoking the definitions of $u$ (\eq{KOpar2}) and  
of the covariant derivative,
\beq\label{Dmu}
D^{ab}_\mu \equiv \delta^{ab} \partial_\mu + g f^{acb} A_\mu^c \ ,  
\eeq
acting on the ghost and anti-ghost fields, one obtains
\beq\label{AmuCAnuCbar}
\langle \left( g f^{abc} A_\mu^b c^c \right) \left( g f^{def} A_\nu^e \overline{c}^f \right) \rangle_k
&=& 
\underbrace{\langle \left( D_\mu^{ac} c^c \right) \left( g f^{def} 
A_\nu^e \overline{c}^f \right) \rangle_k}_{\displaystyle -\delta^{T}_{\mu\nu}(k) \delta^{ad} u(k^2)}
- \underbrace{\langle \ \partial_\mu c^a 
\left( g f^{def} A_\nu^e \overline{c}^f \right) \rangle_k}_{\displaystyle -i k_\mu   
\langle c^a \left( g f^{def} A_\nu^e \overline{c}^f \right) \rangle_k} 
\eeq
which proves that the transversal part of the l.h.s. of \eq{AmuCAnuCbar} is given by \eq{KOpar2},
\beq
\delta_{\mu \mu'}^{T} \ 
\langle 
\left( g f^{abc} A_{\mu'}^b c^c \right) 
\left( g f^{def} A_\nu^e \overline{c}^f \right)
\rangle_k 
\ = \ 
\displaystyle - \delta^{T}_{\mu\nu}(k) \delta^{ad} u(k^2)
\ .
\eeq
while the longitudinal part can be written as follows:
\beq\label{KOpar0}
k_\mu
\langle 
 \left( g f^{abc} A_{\mu}^b c^c \right) 
\left( g f^{def} A_\nu^e \overline{c}^f \right)
\rangle_k 
& = &
i k^2 \langle c^a \left( g f^{def} A_\nu^e \overline{c}^f \right) \rangle_k \nonumber \\
& = & i \underbrace{k^2 \langle c^a \overline{c}^{a'} \rangle}_{\displaystyle  \delta^{aa'} F(k^2)}  
\langle c^{a'} \left( g f^{def} A_\nu^e 
\overline{c}^f \right) 
\rangle_k^{\rm 1PI} \ ;
\eeq
where $F(k^2)$ is the bare ghost propagator dressing function and 
1PI notes the one-particle irreducible contribution to the v.e.v. obtained through the 
amputation of the external ghost leg. Let us then define the function $w(k^2)$ in order 
to parametrize this longitudinal contribution to \eq{AmuCAnuCbar} through
\beq\label{KOpar1}
\langle c^{a} \left( g f^{def} A_\nu^e \overline{c}^f \right) 
\rangle_k^{\rm 1PI}
 =  
i \delta^{ad} \ k_\nu \left( u(k^2) + w(k^2) \right) \ .
\eeq
This definition is equivalent to the one given in terms of diagrams in 
ref.~\cite{Kondo:2009ug,Kondo:2009gc}. Note also that $w(0)$ was
taken to be 0 in the seminal work by Kugo and Ojima.

Taking together eqs.~(\ref{AmuCAnuCbar}) and (\ref{KOpar1})  one  gets:
\beq
\langle \left( g f^{abc} A_\mu^b c^c \right) \left( g f^{def} A_\nu^e \overline{c}^f \right) \rangle_k
\ = - \
\delta^{ad} 
\left( \delta^{T}_{\mu \nu} u(k^2) + F(k^2) \frac{k_\mu k_\nu}{k^2} \ \left( u(k^2) + w(k^2) \right) 
\right)
\ ,
\eeq
and for the v.e.v of the horizon function~\cite{Kondo:2009ug,Kondo:2009gc}:
\beq\label{Eq1}
 \frac{\langle h(0) \rangle_{k=0}}{D(N^2-1)} &=& 
-\frac{1}{D(N^2-1)} \ \langle \ \left( g f^{abc} A_\mu^b \ c^c\right)   
\left( g f^{afe} A_\mu^f \ \overline{c}^e \right) \ \rangle_{k^2 \to 0} \nonumber \\
&=&
- \frac 1 D \left[ (D-1) u(0) + F(0) \left( u(0) + w(0) \right) \ \right] \ .
\eeq 
From now on we make explicit the dependence on the cut-off, $\Lambda$, of all
the bare quantities, which  generally will diverge in the infinite cut-off
limit\footnote{It is well-known that the ghost dressing  function diverges
logarithmically at infinite cut-off in the UV momentum domain.}.  

Kugo has shown in refs.~\cite{Kugo:1995km,Kugo:1979gm} that the Landau gauge condition, $\partial_\mu A_\mu = 0$, 
can be exploited to give:
\beq\label{Eq2}
\left( 1 + u(0,\Lambda) + w(0,\Lambda) \right) F(0,\Lambda) = 1 \ .
\eeq
This result can also  be easily derived from the ghost-propagator Dyson-Schwinger equation  
which, in Landau gauge, can be written as:
\beq
\frac 1 {F(k^2)} \ = \ \frac{\delta^{ab}}{k^2 (N^2-1)} \langle c^a \overline{c}^b \rangle^{-1} 
&=& 
1 - i k_\mu \langle c^a \left( g A_\mu^e f^{def} \overline{c}^f \right) \rangle^{1PI} 
\frac{\delta^{ab}}{k^2 (N^2-1)} \nonumber \\
&=& 1 + u(k^2,\Lambda) + w(k^2,\Lambda)
\eeq
Then, 
the two equations (\ref{Eq1}) and (\ref{Eq2}) can be combined to obtain, {\it without any 
hypothesis about $u$ and $w$}, 
\begin{flushleft}
\begin{align}
u(0,\Lambda) &= \frac{F(0,\Lambda)-1 }{D-1} \ -  \frac{D}{D-1} 
\left[\frac{\langle h(0) \rangle_{k=0}}{D(N^2-1)} \right]
 \notag \\
w(0,\Lambda) &= -1 - u(0,\Lambda) \ + \ \frac 1 {F(0,\Lambda)}\label{sols} \\
 &= - \frac{F(0,\Lambda)+(D-2)}{D-1} 
\ + \ \frac 1 {F(0,\Lambda)} \ + \frac{D}{D-1} 
\left[\frac{\langle h(0) \rangle_{k=0}}{D(N^2-1)} \right]\notag
\end{align}
\end{flushleft}
as general solutions for the Kugo-Ojima parameters, $u(0,\Lambda)$, and
$w(0,\Lambda)$ in  terms of the bare ghost dressing function at vanishing
momentum. {\it If, in addition, we assume that the gap equation~\eq{horizon1}} is satisfied the square bracket 
in \eq{sols} is equal to 1, independently of the cut-off and we get

\beq\label{solgap}\begin{aligned}
u(0,\Lambda) &= \frac{F(0,\Lambda)-1-D }{D-1} \ 
\\
w(0,\Lambda) &= -1 - u(0,\Lambda) \ + \ \frac 1 {F(0,\Lambda)}
 = - \frac{F(0,\Lambda)-2}{D-1} + \ \frac 1 {F(0,\Lambda)}\end{aligned}
\eeq

\begin{figure}
\begin{center}
\includegraphics[width=10cm]{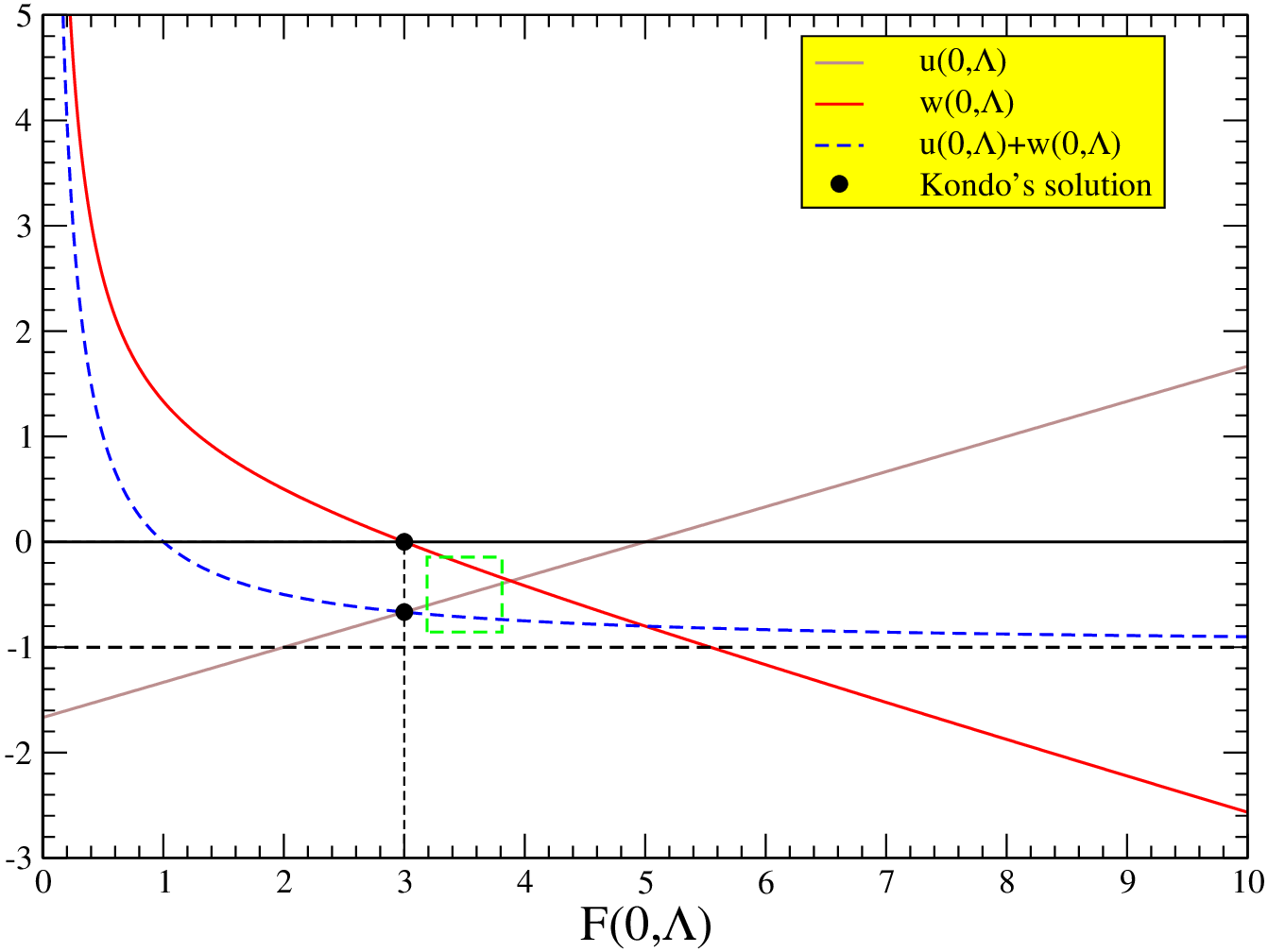}
\end{center}
\caption{{\small The solutions for $u(0,\Lambda)$ and $w(0,\Lambda)$ given by 
\eq{solgap} plotted as a function 
of $F(0,\Lambda)$. This plot can be understood as a function of 
$\widetilde{Z}_3$ for a given non zero value of $F_R(0,\mu^2)$. Then the 
infinite cut-off limit is the limit at infinity of the horizontal axis.  The
particular solution proposed in ref.~\cite{Kondo:2009ug,Kondo:2009gc} (black
circles), obtained  by imposing $w(0,\Lambda)=0$, corresponds to the
intersection of $u+w$ and $u$. It cannot hold when $\widetilde{Z}_3\to \infty$. 
 The current  lattice solutions for the bare ghost dressing functions at
vanishing momentum lie inside the green  dotted square (see
fig.~\ref{fig-ghosts}). This apparent approximate agreement is misleading and
due to the moderate cut-off value on the lattices.}}
\label{kondo-plot}
\end{figure}

In fig.~\ref{kondo-plot}, the  solutions of \eq{solgap} are plotted as 
functions of $F(0,\Lambda)$. It is obvious from \eq{solgap}  (see also
fig.~\ref{kondo-plot}) that, had we required $w(0,\Lambda)=0$, the solution
proposed  in ref~\cite{Kondo:2009ug,Kondo:2009gc} would emerge:
$u(0,\Lambda)=-2/3$ and $F(0,\Lambda)=3$, for $D=4$. However we shall present in the 
next subsection the arguments which lead us  to believe
that no solution implying a cut-off independent and finite value\footnote{This finite value is independent of the number of colors and, 
provided that $w(0,\Lambda)=0$, should be the same whichever regularized
Yang-Mills action we use  (any lattice action, for instance) or even including
any non-zero number of quark flavours for  the action.} for $F(0)$ can be accepted.
\subsection{Constraints from renormalisability}

In this section we assume the validity of the  relation (\ref{horizon1})
Let us start from the basic equation
\beq\label{cutoff}
F(0,\Lambda)
\ = \ 
\widetilde{Z}_3(\mu^2,\Lambda) 
\left( F_R(0,\mu^2) \ + \ {\cal O}\left(\frac 1 {\Lambda^n} \right) \right) \ ,
\eeq
where $F_R$ is the renormalised dressing function in the infinite cut-off limit and $n$ some positive number.
For any quantity which is {\em multiplicatively} renormalisable, 
a field-theory non-perturbative renormalisation scheme (in particular, those applied in lattice 
field theory) 
implies a relation of this kind where the crucial point is that 
{\em the cut-off dependence is an inverse power of the cut-off and cannot 
behave like some power of the cut-off's logarithm}
\cite{reisz1989,Luscher:1998pe}.   
Now, we know  from perturbation theory how $\widetilde{Z}_3(\mu^2,\Lambda)$
depends on the cut-off; 
choosing a fixed value $\Lambda_0$  and sending $\Lambda \to \infty$, 
 $\widetilde{Z}_3$ diverges 
logarithmically in the infinite cut-off limit, regardless of the renormalisation procedure :
\beq\label{cutoffdep}
\frac{\widetilde{Z}_3(\mu^2,\Lambda)}{\widetilde{Z}_3(\mu^2,\Lambda_0)} \ = \  
\left( \frac{\log{(\Lambda/\Lambda_{\rm QCD})}}{\log{(\Lambda_0/\Lambda_{\rm QCD})}} \right)^{9/44} 
\ \left( 1+ \ {\cal O}\left( \alpha\right) \right)\ ,
\eeq
Although this behaviour is quite general, the specific value $\frac{9}{44}$ of the exponent is valid only in the case $N = 3$, $N_f=0$.
 This provides us with two main objections for a finite bare ghost dressing 
function: 

\begin{itemize}
\item A finite value of $F(0,\Lambda)$ requires, through \eq{cutoff}, that   $F_R(0,\mu^2)$ be zero; taking into account the fact that the subdominant terms, 
which vanish as $\Lambda \to \infty$, are supposed not to be logarithmic 
contributions but, at least, of the order of $1/\Lambda$ we are forced to conclude that
zero is the {\em only} allowed finite value that the bare ghost dressing function can hit for \eq{cutoff} to be consistently satisfied.

\item We can  apply \eq{asymp1}
 which will be discussed later and implies at small $q^2$ a 
 cut-off independent factor, decreasing with $q^2$, which multiplies
  $F(0,\Lambda)$ and as well 
 $F_R(0,\mu^2)$. Recalling that the path integration has been limited by the Zwanziger procedure to a a region in which 
 the Faddeev-Popov operator is  positive , we see that if $F_R(0,\mu^2)=0$, $F_R$ can   only assume the value 0 throughout some 
 range of $q^2$, which sounds weird. Even more, there are numerical evidences 
 that the ghost dressing function $F(q^2,\Lambda)$ decreases for all $q^2$. 
 Then  $F_R(q^2,\mu^2)=0$  should hold for any $q^2$. 

\end{itemize}

Therefore, the only way out we see is
 that the bare ghost dressing function diverges 
logarithmically in the infinite cut-off limit and that  multiplication by $\widetilde{Z}_3^{(-1)}$   provides 
 a strictly positive renormalised value. Then, \eq{solgap} can be rewritten as:
\begin{flushleft}\begin{align}\label{sols2}
F_R(0,\mu^2) &= \widetilde{Z}_3^{-1}(\mu^2,\Lambda) \left( \ (D-1) u(0,\Lambda) + D + 1 \ \right)  
\notag \\
&=  (D-1) \left[ \widetilde{Z}_3^{-1}(\mu^2,\Lambda) u(0,\Lambda) + {\cal O}\left( \frac 1 {\log\Lambda} \right)
\right]  \\
u(0,\Lambda) + w(0,\Lambda) &= -1 + {\cal O}\left( \frac 1 {\log\Lambda} \right) \ .\notag
\end{align}\end{flushleft}
An important consequence of the first of equations~(\ref{sols2}) is that the function 
$u(0,\Lambda)$ cannot be  multiplicatively renormalised. The fact that, when multiplied
by $Z_3^{-1}$, it gives a finite result in the infinite cut-off limit does not suffice.
As we have already recalled, it has to differ from its asymptotic value by
inverse powers of the cut-off, which is obviously wrong in \eqs{sols2} where 
$\widetilde{Z}_3^{-1} u(0,\Lambda)$ only converges
up to inverse powers of the {\it logarithm of the cut-off}. The same is true for
$w(0)$. This is not surprising
since  they are defined by the insertion of the composite  operators shown in
eqs.~(\ref{KOpar1},\ref{KOpar2}).

Only the combination  $1+u(0,\Lambda)+w(0,\Lambda)$ vanishes 
logarithmically as $\Lambda \to \infty$ so that,
\beq\label{uwR}
\widetilde{Z}_3(\mu^2,\Lambda) \left( 1 + u(0,\Lambda) + w(0,\Lambda) \right) 
\ = \ 
\frac 1 {F_R(0,\mu^2)} + {\cal O}\left(\frac 1 \Lambda \right) \ ,
\eeq
while both $u(0,\Lambda)$ and $w(0,\Lambda)$ diverge but {their divergences
cancel} in 
\eq{uwR}. Thus, as done in \cite{Aguilar:2009pp,Kondo:2009gc}, 
one can consider $1+u+w$ to be renormalised\footnote{Renormalised  in the sense
that the (logarithmically vanishing) cut-off dependence can be killed at a given
renormalisation  point up to vanishing powers of the cut-off} by
$\widetilde{Z}_3$. However, let us repeat,  $1+u$ and $w$ cannot  be separately renormalised and
$w(0,\Lambda)=0$ cannot be accepted since in conjunction with equations (\ref{horizon1}) and (\ref{sols}) 
 {it provides a finite $u(0,\Lambda)\,=\,-2/3$ and a finite $F(0,
\Lambda)=\,\,3$  which we have shown above to be forbidden}.


\Section{Collecting and extrapolating bare ghost lattice data}\label{revisiting}
\label{sec3}

Lattice simulations first provide us with estimates for bare quantities 
(correlation functions) in the lattice regularization scheme, 
where the role of the regularization cut-off is played by the 
inverse of the lattice spacing, $a^{-1}$ .
In present simulations   $a^{-1}$ is moderate, 
ranging from $\sim 1\, $GeV$^{-1}$ for $\beta=5.7$
up to  $\sim 3.5\, $GeV$^{-1}$ for $\beta=6.4$. 
Those bare quantities should be further renormalised by applying MOM-like 
schemes. That this   multiplicative renormalisation procedure  works  has been proven by Reisz in ref. \cite{reisz1989}; 
the remaining corrections due to finite spacing (vanishing 
in the continuum limit) are considered to behave as powers of the lattice 
spacing. Those renormalised quantities are usually the reliable result 
of the simulations and the ones directly connected with physical predictions.
On the contrary, the  recent work~\cite{Kondo:2009ug,Kondo:2009gc} we  discussed above 
supplies a prediction for a bare quantity: the bare ghost propagator dressing function. Therefore, the 
non-renormalised lattice estimates for this dressing function deserve by  themselves
a particular interest, as far as they could allow us to test that prediction.

In the last few years, many works have been devoted, at least partially, 
to the lattice computation of the ghost propagator. 
They mainly follow ref.~\cite{ZWA94} in writing the Faddev-Popov operator 
as a lattice divergence:
\begin{align}
\label{eq:FP2a}
M(U) = -\frac{1}{N} \nabla\cdot \widetilde{D}(U)
\end{align}
where the operator $\widetilde{D}$ reads
\begin{align}
\label{eq:FP2b}
\widetilde{D}_{\mu}(U)\eta(x) 
   &= \frac{1}{2}\left(U_{\mu}(x)\eta(x+\hat{\mu}) -\eta(x)U_{\mu}(x)
    + \eta(x+\hat{\mu})U^{\dagger}_{\mu}(x)
    - U^{\dagger}_{\mu}(x)\eta(x) \right)
\end{align}
Those definitions, complemented with conversion routines between the Lie algebra and the Lie group, allow for a very efficient lattice
implementation. Some details about the procedure for the inversion of 
the Faddeev-Popov operator and some results will be found 
in \cite{Boucaud:2005gg}.

\begin{figure}
\begin{center}
\includegraphics[width=12cm]{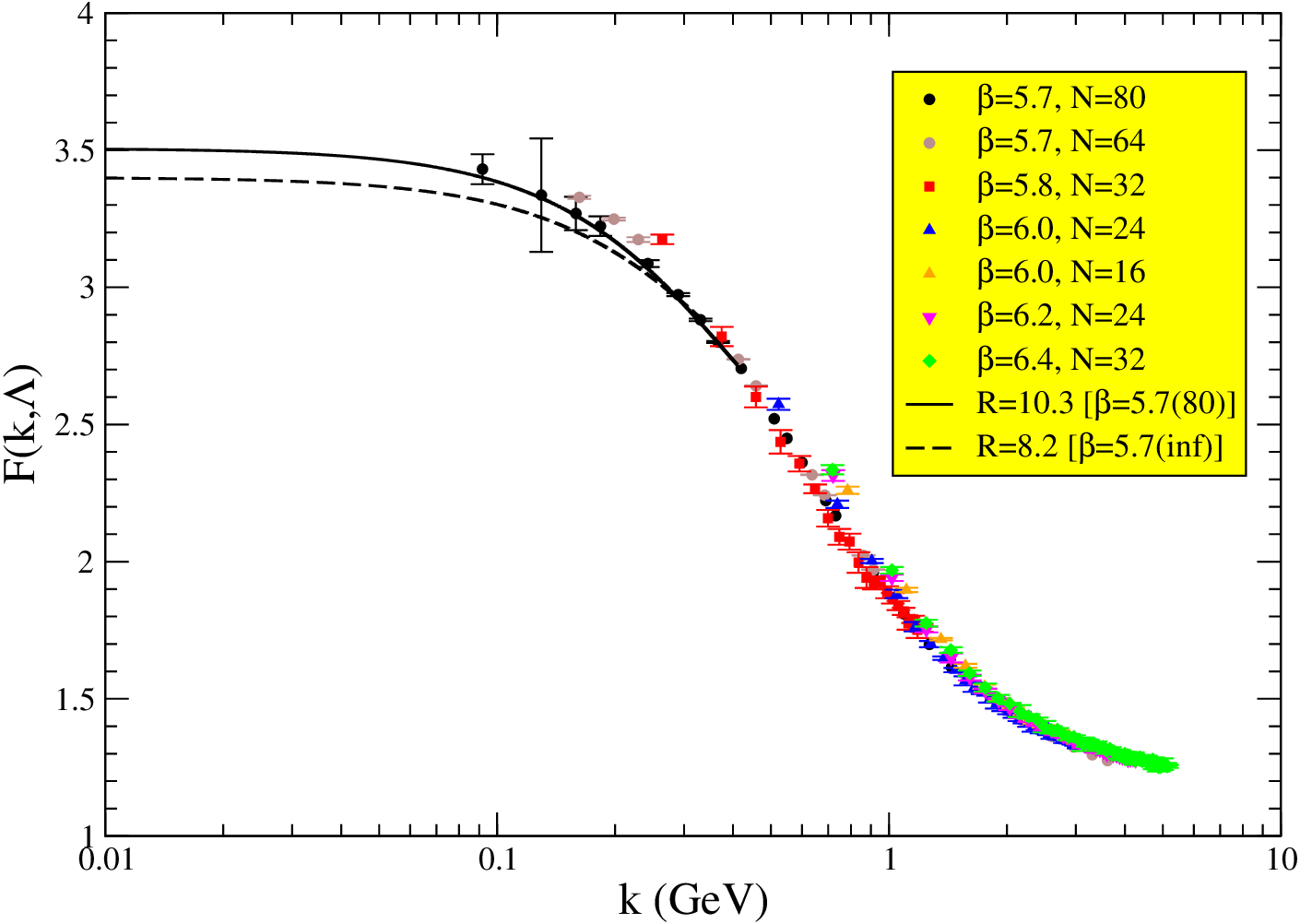}
\end{center}
\caption{{\small Bare ghost dressing function estimated from different lattice data 
sets. The data for the two larger lattice volumes are taken from 
ref.~\cite{Bogolubsky:2009dc} and the others from refs.~\cite{Boucaud:2005gg,Boucaud:2008ji}. 
the solid line is for the best fit with the small-momentum expansion in \eq{asymp1} with 
$R(\beta=5.7(80^4))$ and the dashed one stands for the best fit with $R(\beta=5.7,\infty)$, 
both computed as explained in the text.}}
\label{fig-ghosts}
\end{figure}

The gauge fixing, in particular for Landau gauge, is 
a more delicate issue. A minimization of the functional
\begin{align}
  \label{eq:landau}
  F_{U}[g] =\text{Re}\sum_{x}\sum_{\mu}
  \left(1-\frac{1}{N}g(x)U_{\mu}(x)g^{\dagger}(x+\hat{\mu})\right)
\end{align}
can be achieved by the use of some algorithm driving the gauge
configuration to a local minimum of $F_{U}[g]$. The gauge configurations obtained in this way 
will lie in the first Gribov region but, in general, they do not reach the 
fundamental modular region defined as the set of
{\em absolute} minima of $F_{U}[g]$ on all gauge orbits. A ``best-copy'' algorithm 
(basically consisting in choosing the gauge configuration providing the lowest 
minimum after several minimizations) has also  been used as well as  a 
procedure   that essentially consists in a simulated annealing technique and is claimed 
to reach gauge-functional values closer to the global minima than the standard 
approach (see for instance~\cite{Bogolubsky:2009dc,Bogolubsky:2007ud} and references therein). 
Figure ~\ref{fig-ghosts} presents together results collected from ref.~\cite{Bogolubsky:2009dc} (for very 
big lattice-volume simulations with the simulated annealing gauge-fixing) and data 
from ref.~\cite{Boucaud:2005gg,Boucaud:2005ce,Boucaud:2008ji} (obtained using the standard gauge-fixing); it shows only a weak dependence in the cut-off
$a^{-1}$. This is not surprising since one knows from \eq{cutoffdep}  that it should behave 
at leading log as $\beta^{9/44}$ which gives no larger effect than 2.5 \% on the whole
range of $\beta$'s.  

\begin{figure}
\begin{center}
\includegraphics[width=12cm]{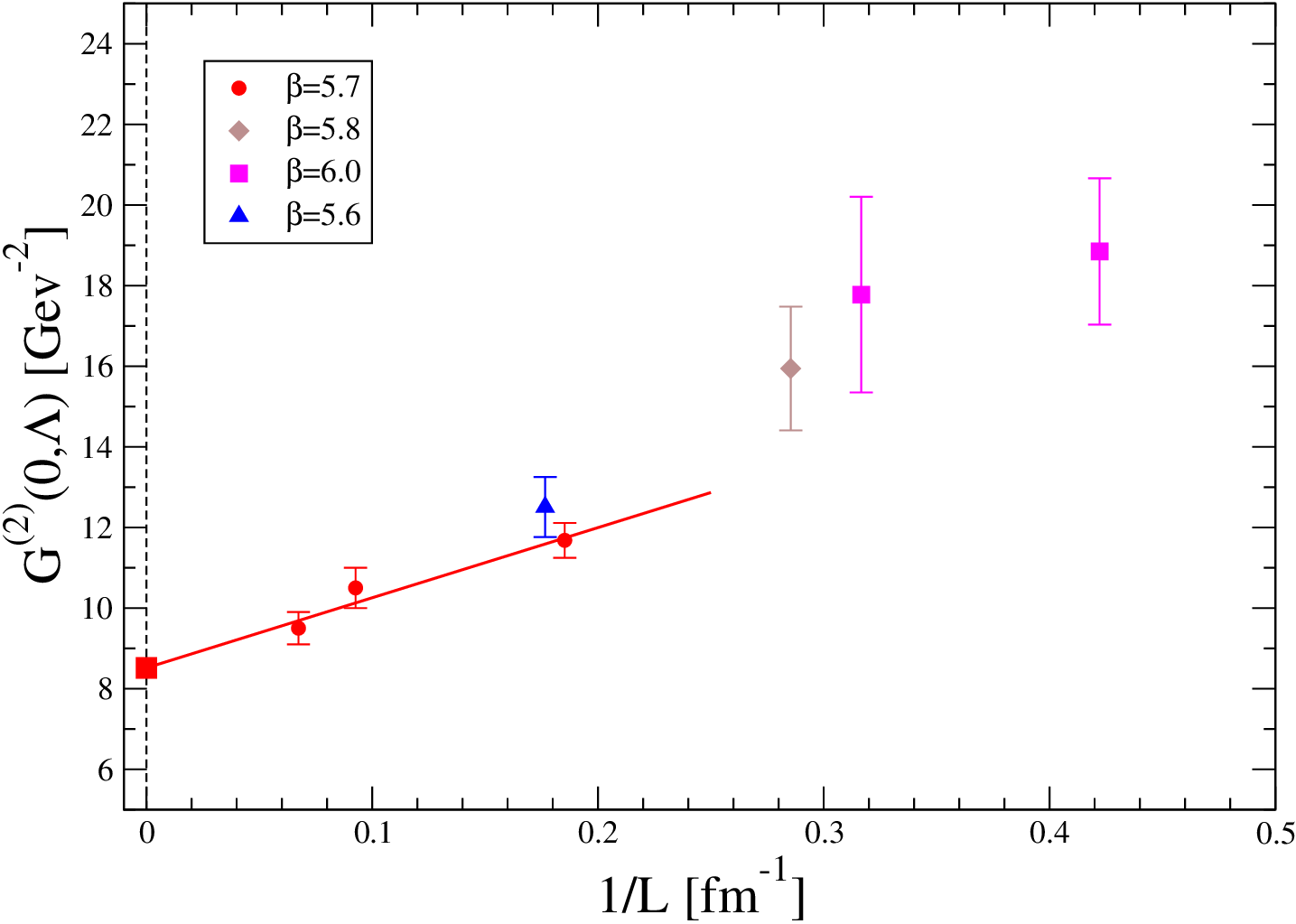}
\end{center}
\caption{{\small Bare zero-momentum gluon propagator estimated from 
different lattice data sets plotted in terms of the inverse of the lattice 
size in physical units. The data for the two bigger lattice volumes are taken 
from ref.~\cite{Bogolubsky:2009dc}, the smaller volume at $\beta=5.7$ corresponds 
to ref.~\cite{Boucaud:2003xi} and the others to refs.~\cite{Boucaud:2005gg,Boucaud:2008ji}. 
A linear fit for the 
three data at $\beta=5.7$ to extrapolate at infinite volume is presented as a 
solid red line.}}
\label{fig-G0}
\end{figure}

In order to compare the lattice data with the previous results we need to extrapolate them to zero momentum. To carry this task out we derive now from the   bare ghost propagator 
Dyson-Schwinger equation (GPDSeq) a small-momentum formula the coefficients of which are fitted against the lattice data. The bare GPDSeq can be regularised and evaluated with the help of a subtraction procedure 
at two different momenta $p$ and $k$, 
\beq\label{GPDSeq}
\frac 1 {F(p^2,\Lambda)} - \frac 1 {F(k^2,\Lambda)}
&=&
N g^2(\Lambda) H_1(\Lambda) \int^{q^2 < \Lambda} \frac{d^4q}{(2 \pi)^4} \
\frac{F(q^2,\Lambda)}{q^2} \left( \frac{(k \cdot q)^2}{k^2}-q^2 \right)
\nonumber \\
&\times& 
\left( \frac{G((q-k)^2,\Lambda)}{(q-k)^4} - 
\frac{G((q-p)^2,\Lambda)}{(q-p)^4} \right)
\eeq
as explained in ref.~\cite{Boucaud:2008ji,Boucaud:2008ky}\footnote{In these
references we dealt with renormalised quantities  but everything applies
straightforwardly for bare ones.}. In this equation $N$ is the number of colours, $g(\Lambda)$ the bare, cut-off dependent, coupling constant and  $G$ stands for the gluon propagator dressing function and 
 $H_1$ is one of the form factors for the bare ghost-gluon vertex,
\beq
\widetilde{\Gamma}_\nu^{abc}(-q,k;q-k) = - i g_0 f^{abc} \left( \ q_\nu H_1(q,k) + (q-k)_\nu H_2(q,k) \ \right) \ ,
\label{DefH12}
\eeq
that should be finite {and only weakly dependent on the momenta} by virtue of 
Taylor's non-renormalisation theorem~\cite{Taylor:1971ff}  and that, consequently, 
is  usually assumed to be constant with respect to the momenta.
Such a bare (and cut-off dependent) GPDSeq can be numerically 
solved, as was done in ref.~\cite{Boucaud:2008ji}, with the help of the lattice gluon propagator estimate 
to be inserted in the integral in \eq{GPDSeq}. It is known (cf. \cite{Boucaud:2008ji}) that the solutions can belong to 2 different types : while the generic solution goes to a finite non-zero limit in the infra-red there exists also an exceptional one (for a given value of the coupling) which diverges as $1/\sqrt{k^2}$ in this same limit. The lattice simulations clearly favour the first type and it is implied in the coming discussion that we are in this situation. The only 
unknown ingredient is the (constant) value  for  the
ghost-gluon vertex form factor, $H_1(\Lambda)$. 
In addition, following ref.~\cite{Boucaud:2008ky}, one can derive a small-momentum expansion for its solution
\begin{flushleft}\begin{subequations}\begin{align}
F(q^2,\Lambda) \ =& \ F(0,\Lambda) \left[ 1 +  \frac{N H_1(\Lambda) R(\Lambda)}{16 \pi} q^2 \log(q^2) 
\ + {\cal O}(q^2) \right] \ , \label{asymp1} \\
\mbox{\rm with:} \ \ \ \ R(\Lambda) =& \frac {g^2(\Lambda)} {4\pi} F(0,\Lambda)^2  
\left( \lim_{k \to 0} \frac{G(k^2,\Lambda)}{k^2} \right) 
\ = \ 
\lim_{k \to 0} \frac{\alpha_T(k^2)}{k^2} + {\cal O}(\frac 1 \Lambda) \ \label{asymp2} ;
\end{align}\end{subequations}\end{flushleft}
%
where $\alpha_T$ is the coupling constant defined in the Taylor scheme~\cite{Boucaud:2008gn}. Before turning to exploit this expansion we shall comment briefly on the various parameters it involves.

Insofar as the gluon propagator reaches a finite non-zero value at vanishing momentum (as it appears to be true on the lattice), 
$R(\Lambda)$ takes a finite value which depends on the cut-off by inverse
powers. Thus, the dominant (in relative terms) $q^2$-dependent part of $F$ in the vicinity of 0 (i.e. 
the second term in the bracket of \eq{asymp1}) {\it does not require any renormalisation}, 
\beq
\lim_{\Lambda \to \infty} \frac{F(q^2,\Lambda)} {F(0,\Lambda)}
\ = \
1 + N H_1 \ \frac{\alpha_T(q^2)}{16 \pi} \log(q^2) + O(q^2) \ .
\eeq
This (quasi-)independence of the slope with respect  to the cut-off is of course important to  
ensure  multiplicative renormalisability, since the latter demands that the $q^2$-dependence of 
the bare  and renormalised Green's functions be the same up to negative powers of the cutoff. 
On the contrary the global multiplicative factor $F(0,\Lambda)$ is known to be logarithmically 
divergent with $\Lambda$, which implies that its variation could be appreciable.

Since $R(\Lambda)$ is to be evaluated from lattice estimates of 
zero-momentum gluon and ghost propagators, one expects it to be sensitive to finite-volume artefacts, to which much attention should therefore
be paid. In 
fig.~\ref{fig-G0} one notices the strong dependence of the zero-momentum gluon propagator on the lattice  size, 
which implies an  equally strong dependence for $R(\Lambda)$. When evaluated from the zero-momentum 
estimates for a $80^4$-lattice at $\beta=5.7$ in ref.~\cite{Bogolubsky:2009dc}, 
$R(\beta=5.7)$ takes on the value of $10.3$ (while, for instance, from the data
for a $32^4$-lattice at $\beta=5.8$ in ref.~\cite{Boucaud:2008ji}, one 
would obtain $R(\beta=5.8)\simeq 19$). 
The bare vertex form factor (supposed to be constant) 
was indirectly estimated in ref.~\cite{Boucaud:2008ji} 
for a $32^4$-lattice at $\beta=5.8$ and appeared to be $H_1(\beta=5.8)\simeq 1.2$.  

Provided that finite-size and lattice spacing artefacts can be neglected for 
the bare ghost-gluon vertex, the values of $R$ 
and $H_1$ we have just determined can be used to attempt to describe the ghost dressing function 
at small momenta estimated from the $80^4$ lattice at $\beta=5.7$ 
in ref.~\cite{Bogolubsky:2009dc}. Then the only parameter in 
\eq{asymp1} which remains to be determined by the best fit
is the zero-momentum ghost dressing 
function. Actually, since a value for $F(0,\Lambda)$ is required in computing $R(\Lambda)$  one 
has to proceed  by iterations: the known value of $G^{(2)}(0,\Lambda)$ and an initial guess of $F(0,\Lambda)$ are inserted in
\eq{asymp2} to produce a first estimate of  $R(\Lambda)$. The latter is then used in \eq{asymp1} to perform a new fit of
 the $80^4$-lattice data deprived of 
the few momenta with $p < 4 \pi /L$. It appears actually that, for all the lattice data sets 
plotted in fig.~\ref{fig-ghosts},  only the momenta satisfying his condition 
are affected by sizeable lattice volume artefacts. This produces a new estimate of $F(0)$ and the process is iterated until it eventually converges to
\beq R(\beta=5.7,80)=10.3 ,\qquad F(\beta=5.7,80) = 3.50 \label{Fbare}.\eeq 
 The fit is presented as a solid line in fig.~\ref{fig-ghosts}.

The impact of the finite-volume effect due to the lattice determination of $R(\Lambda)$ 
can be approximatively 
estimated in the following  way. First, the zero-momentum gluon propagator data for the three 
different lattice volumes at $\beta=5.7$ in fig.~\ref{fig-G0} can 
be extrapolated up to infinite volume (we work only with data for the 
same $\beta$, in order avoid any mixing between lattice-spacing  and volume effects). 
This is the starting point to repeat the iterative procedure explained above and one gets 
$R(\beta=5.7,\infty)=8.2$ and $F(\beta=5.7,\infty) = 3.40$. The extrapolation 
is also shown (with a dashed line) in fig.~\ref{fig-ghosts}.

As for the bare Kugo-Ojima parameter $u(0,\Lambda)$, it is estimated to 
be of the order of $-(0.6 - 0.8)$ 
 in ref.~\cite{Furui:2006nm,Sternbeck:2006rd} and very recently in 
ref.~\cite{Aguilar:2009pp} by using 
a mixed approach, analogous to  the  one previously applied to solve \eq{GPDSeq}, 
in which DS equations are solved with the input of a lattice estimate of the 
gluon propagator.

\subsection{Horizon gap equation and lattice QCD}

A few words are in order to compare the two approaches we have considered in this note.

The Gribov-Zwanziger (GZ) approach \eq{GZ} proposes a modification of the standard 
QCD action in order to limit the domain of the path integration to the domain 
within the Gribov horizon, in which the Faddeev-Popov (FP) operator is positive, i.e.
all its eigenvalues are positive.
 
Lattice QCD uses the genuine QCD action. Some algorithm  minimises the functional 
which discretises the functional $\int d^4x A^\mu_a A^\mu_a$. These algorithms
differ, but they all stop at a local minimum. Local minimum means that 
all the second order derivatives are positive, i.e. that the FP operator is
positive. 

Therefore the two approaches share the property     that they limit themselves to the interior of
the Gribov horizon. None of them manages to find the absolute minimum of that
functional, i.e. to stay within the fondamental domain. There is also a 
``thermodynamic'' argument claiming that one stays close to Gribov's horizon
which means in a domain where the eigenvalues of the FP operator should be small.
This also seems to be valid for both approaches. 

Now come the differences. The GZ approach gives some weight to the different
Gribov copies within the Gribov horizon. The lattice algorithms give another
one, which, moreover,  presumably depends on the specific features  of each algorithm: it has been argued that 
stimulated annealing~\cite{Bogolubsky:2009dc,Bogolubsky:2007ud} leads to smaller 
values of the minimised functional. 

On the lattice it is possible in principle to compute the v.e.v. of the horizon function
  $\langle h(0)\rangle$. Nothing imposes that it should be independent on the cut-off and we do not see any reason why it should verify the horizon gap
equation~\eq{horizon1}. This is precisely a place where the differences we just mentionned 
could be visible.  

In order to discuss this situation let us define a factor $\kappa(\Lambda)$
such that 
\beq\label{horizon1P} \langle h(0) \rangle_{k=0} =  \lim_{k \to 0} \frac 1 {V_D}
\int d^Dx \ \langle h(x) \rangle e^{i k \cdot x} \ = \ \kappa(\Lambda)\;  \left(N^2 - 1 \right) \
D \ . 
\eeq 
The value $\kappa(\Lambda)=1$ corresponds to~\eq{horizon1}. \Eq{sols} now
reads 
\beq\label{solbis}
u(0,\Lambda) &=& \frac{F(0,\Lambda)-1 }{D-1} \ -  \frac{D\, \kappa(\Lambda)}{D-1} 
 \nonumber \\
w(0,\Lambda) &=& -1 - u(0,\Lambda) \ + \ \frac 1 {F(0,\Lambda)}\nonumber \\
 &=& - \frac{F(0,\Lambda)+ D-2}{D-1} 
\ + \ \frac 1 {F(0,\Lambda)} \ + \frac{D \, \kappa(\Lambda)}{D-1} 
\eeq

\begin{figure}
\begin{center}
\includegraphics[width=10cm]{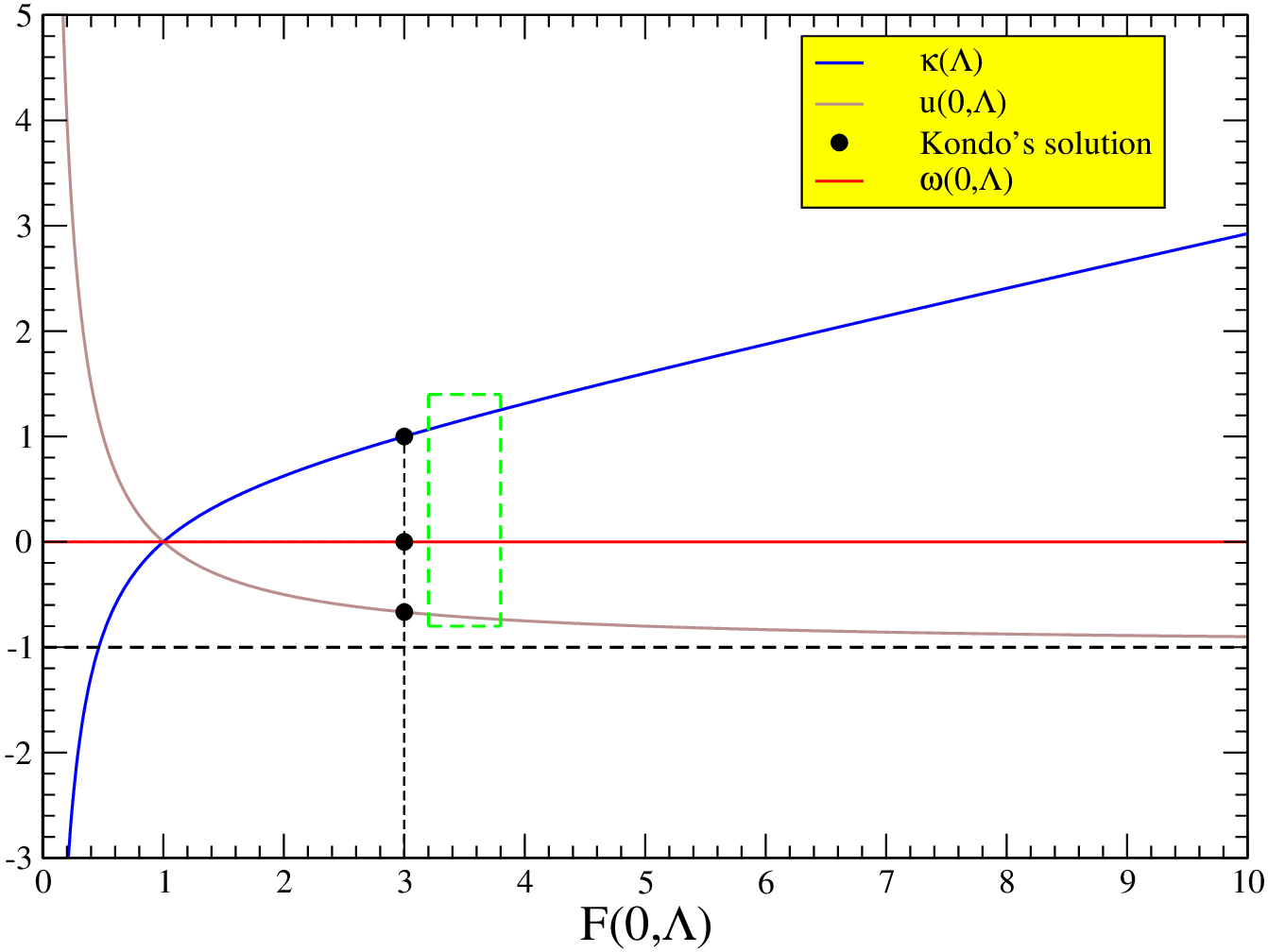}
\end{center}
\caption{{\small The same plot shown in fig.~\ref{kondo-plot} but the gap 
equation is given here by \eq{horizon1P}, the solid blue line being for 
the new factor $\kappa(\Lambda)$ in that equation, and $w(0,\Lambda)$ is required 
to be zero, as explained in the text. 
Again, current lattice estimates lie inside the green dotted square.}}
\label{w0-plot}
\end{figure}

If the gap equation~\eq{horizon1} is  relaxed in this way it becomes possible to 
have $w(0)=0$ as shown in~\cite{Aguilar:2009pp,Grassi:2004yq} in the Landau
background gauge and assumed in~\cite{Kondo:2009ug,Kondo:2009gc}. The solution
becomes
\beq
\kappa(\Lambda) &=& \frac{F(0,\Lambda)}{D} + \frac{D-2}{D}  - \frac{D-1}{D\,F(0,\Lambda)}  
\nonumber \\
u(0,\Lambda) &=& \frac{1}{F(0,\Lambda)} -1 \, .
\eeq
implying that 
$\kappa(\Lambda)\to \infty$ and $u(0,\Lambda)\to -1$ when $\Lambda \to \infty$ 
(see fig.~\ref{w0-plot}).
This does not change our major conclusion that no finite value of $F(0,\Lambda)$
independent of $\Lambda$ is acceptable. In particular it happens that 
$\mathbf {u(0,\Lambda \to \infty)}$ {\bf converges to }$-1$, {\bf
nevertheless Kugo-Ojima's condition 
only emerges at the infinite cut-off limit and thus } $\mathbf{ 0 < F_R(0, \mu^2) < \infty}$. 

 Lattice measurements, which correspond to $\Lambda \sim 1-4 $ GeV, 
give an estimate of 
 $\kappa(\Lambda)$  which, with $u \sim (.6-.8)$  and F(0)$\sim 3.5$, leads to 
 $\kappa(\Lambda) \simeq 1.2-1.3$. However this does not tell how  
 $\kappa(\Lambda)$ depends on  $\Lambda$.


\section{Conclusions}


We have generalized the solution recently proposed by Kondo for the zero-momentum ghost dressing function and the Kugo-Ojima parameter, $u(0,\Lambda)$, by deriving eqs. (\ref{sols}) where both Kugo-Ojima parameters, $u(0,\Lambda)$ and $w(0,\Lambda)$ appear written in terms of $F(0,\Lambda)$ and the horizon function, $\langle h(0,\Lambda)\rangle$, at any finite cut-off. In particular, we have shown that the one relating $u(0,\Lambda)$ and $F(0,\Lambda)$, after applying the gap 
equation, \eq{horizon1}, is close to be verified by lattice estimates for 
$\Lambda = a^{-1} \sim 1 $GeV, but that this
is a pure coincidence due to the small cut-off in lattice calculations.
We have argued that neither
$u$ nor $w$ can be multiplicatively renormalised. Indeed,
from the anomalous dimension of the ghost propagator renormalisation constant 
we conclude that no solution  with a cut-off independent bare $F(0,\Lambda)$ is 
possible. If the gap equation is valid, for $\Lambda \to \infty$, then
 $u(0,\Lambda)\to \infty$ and $w(0,\Lambda)\to -\infty$ such that 
 $u(0,\Lambda)+w(0,\Lambda)\to -1$. If one relaxes the gap 
 equation~\eq{horizon1P}, one can satisfy  $w(0,\Lambda)=0$ with
 $\kappa(\Lambda)\to \infty$ and $u(0,\Lambda)\to -1$. The Kugo-Ojima condition 
 $u(0)=-1$ is asymptotically fulfilled for the bare $u$ while the renormalised 
 ghost dressing function is finite and non zero.  
We have argued that lattice QCD,
notwithstanding some similarity with the Gribov-Zwanziger approach, has no reason to
fulfill Zwanziger's horizon gap equation. We have shown, however,  that this fact will not change
much about our conclusions concerning the ghost propagator. 

Our major conclusion about the ghost propagator is obtained from the joint use of  lattice data 
and of a result stemming from the ghost propagator Dyson-Schwinger equation: this
result consists in a simple and cut-off independent formula for the ghost propagator
dependence at small momentum.  If we choose 1.5 GeV as the renormalisation scale, 
we get from lattice 
\beq
F\,(1.5 \,\mathrm{GeV}) \equiv \widetilde{Z_3} \simeq 1.6 \quad\mathrm{whence}\quad
F_R\,(0,1.5 \,\mathrm{GeV}) \simeq 2.2.
\eeq    
This has of course to be refined particularly regarding finite volume effects which should be considered with
more care. The major point in this note, from the point of view of the renormalisability of 
the theory, is that lattice artefacts behave as  powers of $a$ (in this case $O(a^2)$). 
Of course at $\beta =5.7$ lattice spacing is not  yet small and this leads to a 
significant uncertainty which deserves further study. Any statement from lattice concerning bare quantities has to be taken with great
caution since the very slow logarithmic dependence has chances to escape
numerical observation.

\paragraph{Acknowledgements:} 
It is a pleasure for the authors to acknowledge fruitful discussions with
K-I. Kondo, D. Dudal, A. C. Aguilar, D. Binosi and J. Papavassiliou. This 
work has been partially funded by the spanish research project 
FPA2006-13825.


\end{document}